\documentclass[leqno,a4paper,11pt]{article}
\usepackage{amsmath, amssymb, ascmac, latexsym, amsthm}
\usepackage{color}

\theoremstyle{definition}
\newtheorem{Assu}{Assumption}

\newtheorem{Th}{Theorem}[section]
\newtheorem{Def}{Definition}
\newtheorem{Lem}[Th]{Lemma}
\newtheorem{Prop}[Th]{Proposition}
\newtheorem{Cor}[Th]{Corollary}
\newtheorem{Rem}[Th]{Remark}

\title{Wave front set for solutions to Schr\"odinger equations with time-dependent variable coefficients}
\author{Keiichi Kato, Tetsuya Ogawa, Taisuke Yoneyama}
\date{\today}
\newcommand{\dint}{\displaystyle\int}

\begin{document}
\maketitle
\begin{abstract}
In this paper, we determine the $C^\infty$-type wave front sets of the
solutions to the Schr\"odinger equations with time-dependent variable coefficients and potentials by using the wave packet transform.
We introduce the infinite sum and estimate the perturb terms of the flat Laplacian on $\mathbb{R}^n$,
which cannot be estimated by the method in Kato--Ito \cite{KaI}.
\end{abstract}
%
%
%
\section{Introduction}
In this paper, we investigate the wave front set of the solutions to the following Schr\"odinger equations with time-dependent variable coefficients:
\begin{align}
\label{CP1}
&\begin{cases}
i\partial_tu+ \displaystyle\frac{1}{2}\sum_{j,k=1}^{n} \partial_{x_j} a_{j,k}(t,x)\partial_{x_k}u -V(t,x)u= 0, \quad &(t,x)\in\mathbb{R}\times\mathbb{R}^{n},\\
u(0,x)=u_0(x),\hspace{100pt} &x\in\mathbb{R}^n,
\end{cases}
\end{align}
where the coefficients $a_{j,k}(t,x)$ and the potential $V(t,x)$ satisfy the
following assumptions:\\
\begin{Assu}\label{assu}
For $i,j\geq1$, $a_{j,k}=a_{k,j}$, $a_{j,k}(t,x)$ and $V(t,x)$ are in $C^{\infty}(\mathbb{R}\times\mathbb{R}^n;\mathbb{R})$
and
there exist $\rho>1$ and $C_\alpha>0$
for each $\alpha\in\mathbb{Z}_+^n$ such that
\begin{align}
\label{as1}
|\partial_x^\alpha(a_{j,k}(t,x)-\delta_{j,k})|\leq C_\alpha(1+|x|)^{-\rho-|\alpha|},\quad (t,x)\in\mathbb{R}\times\mathbb{R}^n,\\
\label{as2}
|\partial_x^\alpha V(t,x)|\leq C_\alpha(1+|x|)^{2-\rho-|\alpha|},\quad (t,x)\in\mathbb{R}\times\mathbb{R}^n,
\end{align}
where $\delta_{j,k}$ is the Kronecker delta.
\end{Assu}

In addition, we assume the existence of the propagater of $\eqref{CP1}$.
\begin{Assu}\label{assu2}
There exists a family of unitary operators $(U(t,\tau))_{(t,\tau)\in\mathbb{R}^2}$ in $L^2(\mathbb{R}^n)$  satisfying the following conditions.

\noindent(i) For $f\in L^2(\mathbb{R}^n)$, $U(t,\tau)f$ is strongly continuous function with respect to $t$ and satisfies
\[
U(t,\tau')U(\tau',\tau)=U(t,\tau),\,U(t,t)=I\quad\mbox{for all }t,\tau',\tau\in\mathbb{R},
\]
where $I$ is the identity operator on $L^2(\mathbb{R}^n)$.\\
(ii) For $u_0\in L^2(\mathbb{R}^{n})$, $u(t)=U(t,0)u_0$ is strongly continuously differentiable in $L^2(\mathbb{R}^n)$ with respect to $t$ and satisfies
$\eqref{CP1}$.
\end{Assu}
%
%
In this paper, we write $\mathcal{S}(\mathbb{R}^n)$ and $\mathcal{S}'(\mathbb{R}^n)$
as the Schwartz space and the set of the temperd distributions, respectively.
\begin{Def}[$C^\infty$-type wave front set]\label{wave-front-set}
Let $f\in \mathcal{S}'(\mathbb{R}^n)$ and $(x_0,\xi_0)\in\mathbb{R}^n\times( {\mathbb{R}^n\setminus\{0\})}$.
Then we say that $(x_0,\xi_0)\notin WF(f)$ if and only if
there exist $\chi\in C_0^{\infty}(\mathbb{R}^n)$ with $\chi (x_0) \not= 0$ and a conic neighborhood $\Gamma$ of $\xi_0$ ({\rm i.e.,} $\xi\in\Gamma$ and $c>0$ implies $c\xi\in\Gamma$)
satisfying that
for $N\in\mathbb{N}$ there exists $C_N>0$ such that
$$|\widehat{\chi f}(\xi)| \leq C_N(1+|\xi|)^{-N}$$
for any $\xi\in\Gamma$,
where $\hat{f}(\xi)=\mathcal{F} [f](\xi)\equiv(2\pi)^{-n/2}\int_{\mathbb{R}^n}f(x)e^{-ix\cdot\xi}dx$
denotes the Fourier transform of $f$.
\end{Def}
This wave front set is introduced by L. H\"ormander \cite{Hor},
which is one of the main tools of microlocal analysis in $C^\infty$ category.
G. B. Folland  \cite {Fo} introduced the characterization of the wave front set in terms of
wave packet transform which is introduced by A. C\'ordoba--C. Fefferman \cite{CF}.
%
%
\begin{Def}[Wave packet transform]\label{wpt-def}
Let $\varphi\in\mathcal{S}(\mathbb{R}^n)\setminus\left\{ 0\right\}$,
$f\in\mathcal{S}'(\mathbb{R}^{n})$ and $F(y,\xi)$ be a function on $\mathbb{R}^{n}\times\mathbb{R}^{n}$.
We define the wave packet transform $W_\varphi f(x,\xi)$ of $f$ 
with the wave packet generated by a window $\varphi$ as follows:
\[
W_\varphi f(x,\xi)=\int_{\mathbb{R}^{n}}\overline{\varphi(y-x)}f(y)e^{-iy\xi}dy 
\quad\mbox{for } (x,\xi)\in\mathbb{R}^{n}\times\mathbb{R}^{n}.
\]
\end{Def}
The main statement of this paper is following.
%
%
\begin{Th}\label{main-theorem}
Suppose that Assumption $(A)$ and $(B)$ are satisfied. Let $u_0\in L^2(\mathbb{R}^n)$, $u(t)=U(t,0)u_0$, $\varphi_0(x)=e^{-|x|^2/2}$ and
\begin{align}\label{w-l}
\varphi_\lambda(t,x)
=e^{it\Delta/2}\lambda^{n/4}\varphi_0(\lambda^{1/2} x)
=\mathcal{F}^{-1}[e^{-it|\xi|^2/2}\lambda^{n/4}\mathcal{F}[\varphi_0(\lambda^{1/2} \cdot)](\xi)](x).
\end{align}
Then $(x_0,\xi_0)\notin WF(u(t,\cdot))$ if and only if
there exist a neighborhood  $K$ of $x_0$ and a conic neighborhood $\Gamma$ of $\xi_0$ 
such that for all $N\in\mathbb{N}$, $a\geq1$ and $\varphi_0\in \mathcal{S}(\mathbb{R}^n)\setminus\{0\}$,
there exists $C_{N,a,\varphi_0}>0$ satisfying
\begin{align}\label{m-th}
|W_{\varphi_{\lambda}{(-t)}}u_0 (x(0,t,x,\lambda\xi), \xi(0,t,x,\lambda\xi))|\leq C_{N,a,\varphi_0}\lambda^{-N}
\end{align}
for $\lambda\geq1$ and $(x,\xi)\in K\times\Gamma$ with $a^{-1}\leq|\xi|\leq a$.
Here 
$x(s)=x(s;t,x,\xi)$ and $\xi(s)=\xi(s;t,x,\xi)$ are the solutions of
\begin{align}
\label{h-eq}
\begin{cases}
\dfrac{d}{ds} x(s)=\dfrac{\partial H}{\partial \xi}(s,x(s),\xi(s)),\ x(t)=x, \\[3mm]
\dfrac{d}{ds} \xi(s)=-\dfrac{\partial H}{\partial x}(s,x(s),\xi(s)),\ \xi(t)=\xi,\\
\end{cases}
\end{align}
where $H(t,x,\xi)=-(1/2)\sum_{j,k=1}^{n}a_{j,k}(t,x)\xi_j\xi_k+V(t,x)$.
\end{Th}
%
%
Using the properties of the classical orbits, we obtain the following corollary.
\begin{Cor}\label{Cor2}
Under the same assumpton and notation in Theorem \ref{main-theorem}.
Then $(x_0,\xi_0)\notin WF(u(t,\cdot))$ if and only if
there exist a neighborhood $K$ of $x_0$ and a conic neighborhood $\Gamma$ of $\xi_0$
such that for all $N\in\mathbb{N}$, $a\geq1$ and $\varphi_0\in \mathcal{S}(\mathbb{R}^n)\setminus\{0\}$,
there exists $C_{N,a,\varphi_0}>0$ satisfying
\begin{align}\label{m-th-c}
|W_{\varphi_{\lambda}(-t)}u_0 (x-t\lambda\xi,\lambda\xi)|\leq C_{N,a,\varphi_0}\lambda^{-N}
\end{align}
for $\lambda\geq1$ and $(x,\xi)\in K\times\Gamma$ with $a^{-1}\leq|\xi|\leq a$.
\end{Cor}
\begin{Rem}
In \cite{KaI}, the similar corollary was proven.
However, there is insufficiency of the argument in the proof.
In this paper, we make up for the insufficiency in Section 4.
\end{Rem}
%
%
In the previous studies of the wave front set, G. B. Folland used the basic wave packet as a window.
T. \=Okaji \cite{Oo} obtained the same conclusion with a general window under some conditions.
After that, K. Kato--K. Kobayashi--S. Ito \cite{KKI} removed some restriction of window completely.

The following results of wave front sets for the solutions of the Schr\"odinger equation without time-dependent variable coefficients are well-kwown.
A. Hassell--J. Wunsch \cite{HW} and S. Nakamura \cite{Na} has studied the wave front set for the time-independent case i.e., 
$a_{j,k}(t,x)=a_{j,k}(x)$ and $V(t,x)=V(x)$.
A. Hassell--J. Wunsch investigated the quadratic scattering wave front set which is introduced by Melrose \cite{Me}.
S. Nakamura used the semi-classical way with the pseudo-differential operators.
On the other hand, K. Kato--S. Ito \cite{KaI} has treated the case $a_{j,k}(t,x)=\delta_{j,k}$ with time-dependent sub-quadratic potentials
by using the wave packet transform instead of the pseudo-differential operators,
however, this method cannnot be applied directly for the solution to the equation with general coefficients $a_{j,k}(t,x)$.
Thus we use the series function $\Phi(t)$ defined in \eqref{def-ph} and apply the Gronwall inequality for $\Phi(t)$
and thereby we get the caracterization of the wave front set for the solution to the equation with general coefficients $a_{j,k}(t,x)$.

Our result is different from the previous results by including time-dependent variable coefficients.
In \cite{KaI}, time-dependent potentials are treated but the principal part does not depend on time.
In order to treat time-dependent case, we use the phase space analysis, in particular, we use the wave packet transform.
The idea of the proofs of the main theorems is as follows.
Splitting the part $(1/2)\sum\partial_{x_j} a_{j,k}(t,x)\partial_{x_k}u -V(t,x)u$ into $((1/2)\sum\partial_{x_j} (a_{j,k}-\delta_{j,k})(t,x)\partial_{x_k}u)+((1/2)\Delta u-V(t,x)u)$
and estimating the each term by using the wave packet transform,
we prove the decay estimate with respect to $\lambda$.
The latter part can be estimated by the similar way in \cite{KaI}.
However, the estimate of the former part cannot be applied to the method introduced in \cite{KaI}.
Hence the key of our proof is to estimate the former part.
In order to prove this, we need to estimate the term like $\lambda^2\sum_{|\alpha|=2}\xi^\alpha W_\varphi[(a_{j,k}(t)-\delta_{j,k})u(t)](x,\lambda \xi)$,
which does not appear in \cite{KaI}.
Thus we introduce the series $\Phi(t)$ defined in $\eqref{def-ph}$ and take a window like $\eqref{w-l}$.
This idea can solve the difficulties.
In the proof of Corollary \ref{Cor2}, the new method is used.
Introducing the inverse function of $(x(s),\xi(s))$,
we prove the smallness between $(x(s),\xi(s))$ and $(x-t\xi,\xi)$ and find suitable neighborhoods of $x$ and $\xi$.

\begin{Rem}
Assumption $\mathrm{(B)}$ is satisfied by the Stone theorem under Assumption $\mathrm{(A)}$
if $a_{j,k}$ and $V$ do not depend on $t$ and $\det(a_{j,k})> 0$ for any $x\in\mathbb{R}^n$.
\end{Rem}

In this paper, we use the following notaitons.
$i=\sqrt{-1}$, $n\in\mathbb{N}$, $\partial_{x_j}=\partial/\partial_{x_j}$, $\partial_t=\partial/\partial t$,
$\Delta=\sum_{j=1}^{n}\partial x_j^2$ and $\int =\int_{\mathbb{R}^n}$. 
We denote $B_r(x)$ as a open ball $\{y\in\mathbb{R}^n\,|\,|y-x|<r\}$.
We often write $W_\varphi u(t,x,\xi)$ as $W_\varphi [u(t)](x,\xi)$.
$e_j=(0,\ldots,0,\overset{j}{\check{1}},0,\ldots,0)$.
For a vector $x\in\mathbb{R}^n$, $\langle x\rangle=\sqrt{1+|x|^2}$.
For a multi-index $\alpha=(\alpha_1, \cdots, \alpha_n),\,\beta\in\mathbb{Z}_+^n$,
$|\alpha|=\alpha_1+\cdots+\alpha_n$,
$x^{\alpha}=x_1^{\alpha_1}\cdots x_n^{\alpha_n}$, $\partial_{x}^{\alpha}=\partial_{x_1}^{\alpha_1}\cdots\partial_{x_n}^{\alpha_n}$ and
$p^{(\alpha)}_{(\beta)}(x)=x^\beta\partial_x^\alpha p(x)$ for $p\in C^\infty(\mathbb{R}^n)$.
The plan of the paper is as follows.
In section 2, we shall study the properties of the classical orbits.
In section 3, we give a proof of Theorem \ref{main-theorem}.
In section 4, we give a proof of Corollary \ref{Cor2}.

%
%

\section{Property of classical orbits}\label{co}

In this section,  we study the property of the classical orbits which are solutions to $\eqref{h-eq}$.

The following property is used in the proof of the Theorem \ref{main-theorem}.
\begin{Prop}\label{c-o-prop}
Suppose that Assumption $(A)$ is satisfied.
Then the solutions $x(s)=x(s;t_0,x,\xi)$ and $\xi(s)=\xi(s;t_0,x,\xi)$
of $\eqref{h-eq}$ exist uniquely and are $C^\infty$ class in term of $(x,\xi)$
and their derivatives are $C^1$ class in term of $s$.
Moreover, there exists a constant $\lambda_0\geq1$  such  that
\begin{align}\label{c-o-p-1}
\frac{1}{2}\lambda\langle s-t_0 \rangle|\xi|\leq1+|x(s;t_0,x,\lambda\xi)|\leq 2 \lambda \langle s-t_0 \rangle|\xi|,
\end{align}
and
\begin{align}\label{c-o-p-2}
\frac{1}{2}\lambda|\xi|\leq|\xi(s;t_0,x,\lambda\xi)|\leq 2\lambda|\xi|.
\end{align}
for $\lambda\geq\lambda_0$, $|s-t_0|\leq|t_0|$ and $(x,\xi)\in K\times\Gamma$ with $a^{-1}\leq|\xi|\leq a$.
\end{Prop}
\begin{proof}
We define $x_N(s)$ and $\xi_N(s)$ as follows:
\begin{align*}
\begin{cases}
x_0(s)=x+\lambda(s-t_0)\xi,\\
\xi_0(s)=\lambda\xi
\end{cases}
\end{align*}
and for $N\geq1$
\begin{align*}
\begin{cases}
x_N(s)=x+\dint_{t_0}^s\xi_N(\theta)d\theta+\dint_{t_0}^s(A(\theta,x_{N-1}(\theta))-E)\xi_{N-1}(\theta)d\theta,\\
\xi_{N}(s)=\lambda\xi-\displaystyle\frac{1}{2}\sum_{j,k=1}^n\dint_{t_0}^s\nabla_x a_{j,k}(\theta,x_{N-1}(\theta))\xi^j_{N-1}(\theta)\xi^k_{N-1,}(\theta)d\theta
+\dint_{t_0}^s\nabla_x V(\theta,x_{N-1}(\theta))d\theta,
\end{cases}
\end{align*}
where $A(t,x)=(a_{j,k}(t,x))$, $E$ denotes the $n\times n$ identity matrix and
$\xi_N(s)={}^t(\xi^1_N,\ldots,\xi^n_N)$.

The existence of the solutions of $\eqref{h-eq}$ follows from the Picard iteration method.
Thus $x(s)$ and $\xi(s)$ are obtained by $\lim_{N\to\infty}x_N(s)=x(s), \lim_{N\to\infty}\xi_N(s)=\xi(s)$.
We only show (\ref{c-o-p-1}) and (\ref{c-o-p-2}) with $x(s)=x_N(s)$ and $\xi(s)=\xi_N(s)$
by induction.

Clearly, (\ref{c-o-p-1}) and (\ref{c-o-p-2}) hold for $N=0$.
Assuming that (\ref{c-o-p-1}) and (\ref{c-o-p-2}) hold with $x(s)=x_l(s)$ and $\xi(s)=\xi_l(s)$,
we have
\begin{align*}
\left|\xi_{l+1}(s)\right|&\leq\lambda|\xi|
+\dfrac{1}{2}\sum_{j,k=1}^n\left|\dint_{t_0}^s\nabla_x a_{j,k}(\theta,x_{l}(\theta))\xi^j_{l}(\theta)\xi^k_{l}(\theta)d\theta\right|
+\left|\dint_{t_0}^s\nabla_x V(\theta,x_{l}(\theta))d\theta\right|\\
&\leq\lambda|\xi|
+\dfrac{1}{2}\sum_{j,k=1}^n\dint_s^{t_0}C(1+|x_l(\theta)|)^{-\rho-1}|\xi^j_{l}(\theta)||\xi^k_{l}(\theta)|d\theta
+\dint_s^{t_0} C(1+|x_{l}(\theta)|)^{-\rho+1}d\theta\\
&\leq\lambda|\xi|
+ C\dint_s^{t_0}(1+|x_l(\theta)|)^{-\rho-1}\left(\frac{1+|x_l(\theta)|}{\langle \theta-t_0 \rangle}\right)^2d\theta
+C\times |t_0|\\
&\leq \lambda|\xi|\left(1+\frac{C''}{\lambda|\xi|}\right).
\end{align*}
Here we use $\langle s-t_0 \rangle|\xi_l(s)|\leq 2\langle s-t_0 \rangle\lambda|\xi|\leq 4(|x_l(s)|+1)$.
Hence, taking $\lambda_0$ sufficiently large, we have $|\xi_{l+1}(s)|\leq 2\lambda|\xi|$
for $\lambda\geq\lambda_0$.
$|\xi_{l+1}(s)| \geq \lambda|\xi|/2$ can be shown in the same way.

On the other hand, we have
\begin{align*}
1+|x_{l+1}(s)|&=1+\left|x+\dint_{t_0}^s\xi_{l+1}(\theta)d\theta+\dint_{t_0}^s(A(\theta,x_{l}(\theta))-E)\xi_{l}(\theta)d\theta\right|\\
&\geq1+\left|\dint_{t_0}^s\xi_{l+1}(\theta)d\theta\right|
-|x|-\left|\dint_{t_0}^s(A(\theta,x_{l}(\theta))-E)\xi_{l}(\theta)d\theta\right|\\
&\geq1+\lambda|s-t_0||\xi|-|x|-\displaystyle\frac{1}{2}
\sum_{j,k=1}^n\left|\dint_{t_0}^s\dint_{t_0}^{s_1}\nabla_x a_{j,k}(\theta,x_l(\theta))\xi^j_{l}(\theta)\xi^k_{l}(\theta)d\theta ds_1\right|\\
&\quad-\displaystyle\left|\dint_{t_0}^s\nabla_x V(\theta,x_{l}(\theta))d\theta\right|
-\left|\dint_{t_0}^s(A(\theta,x_{l}(\theta))-E)\xi_{l}(\theta)d\theta\right|\\
&\geq1+\lambda|s-t_0||\xi|-|x|-C(s-t_0)^2-C|s-t_0|\\
&\geq \lambda\langle s-t_0 \rangle|\xi|\Big(1-\frac{C}{\lambda}\Big).
\end{align*}
Thus we have $|x_{l+1}(s)|\geq \lambda\langle s-t_0 \rangle|\xi|/2$ for large $\lambda$.
Similarly, $|x_{l+1}(s)|\leq 2\lambda\langle s-t_0 \rangle|\xi|$ holds.
Hence (\ref{c-o-p-1}), (\ref{c-o-p-2}) are obtained with $x(s)=x_{l+1}(s)$ and $\xi(s)=\xi_{l+1}(s)$.
\end{proof}
We can obtain the following proposition by the similar way in the proof of Proposition \ref{c-o-prop}.
\begin{Prop}\label{c-l-1}
Suppose that Assumption $(A)$ is satisfied.
Then the solutions $y(s)=y(s;t,y,\eta)$ and $\eta(s)=\eta(s;t,y,\eta)$
of 
\begin{align}\label{h-eq2}
\begin{cases}
\dfrac{d}{ds} y(s)=\dfrac{\partial H}{\partial \xi}(s,y(s),\eta(s)),\ y(0)=y-t\eta, \\[3mm]
\dfrac{d}{ds} \eta(s)=-\dfrac{\partial H}{\partial x}(s,y(s),\eta(s)),\ \eta(0)=\eta.\\
\end{cases}
\end{align}
exist uniquely and are $C^\infty$ class in term of $(x,\xi)$
and their derivatives are $C^1$ class in term of $s$.
Moreover, there exists a constant $C>0$  such  that
\begin{align}\label{l-decay}
|y(t;t,y,\lambda\eta) -y|&\leq C\lambda^{-\rho+1}\\
|\eta(t;t,y,\lambda\eta) -\lambda\eta|&\leq C\lambda^{-\rho+1}
\end{align}
for $\lambda\geq1$, $|s-t_0|\leq|t_0|$ and $(x,\xi)\in K\times\Gamma$ with $a^{-1}\leq|\xi|\leq a$.
\end{Prop}

The relation between $(x(s),\xi(s))$ and $(y(s),\eta(s))$ is following.
\begin{Lem}\label{c-l-2}
Suppose that Assumption $(A)$ is satisfied.
Let $(x(s),\xi(s))$ and $(y(s),\eta(s))$ be solutions to $\eqref{h-eq}$ and $\eqref{h-eq2}$, respectively.
Then the following identities hold:
\begin{align}\label{l-decay}
\begin{cases}
x(0;t,y(t;t,y,\eta),\eta(t;t,y,\eta))&=y-t\eta\\
\eta(0;t,y(t;t,y,\eta),\eta(t;t,y,\eta))&= \eta
\end{cases}
\end{align}
for $\lambda\geq1$, $|s-t_0|\leq|t_0|$ and $(x,\xi)\in K\times\Gamma$ with $a^{-1}\leq|\xi|\leq a$.
\end{Lem}
\begin{proof}
By the uniqueness of the solutions, we have for $s\leq t$ and $(x,\xi)\in K\times\Gamma$ with $a^{-1}\leq|\xi|\leq a$
\begin{align*}
\begin{cases}
x(s;t,y(t;t,y,\eta),\eta(t;t,y,\eta))=y(s;t,y,\eta),\\
\xi(s;t,y(t;t,y,\eta),\eta(t;t,y,\eta))=\eta(s;t,y,\eta),
\end{cases}
\end{align*}
which implies $\eqref{l-decay}$.
\end{proof}

%
%
\section{Proof of Theorem\ref{main-theorem}}
In this section, we give a proof of Theorem \ref{main-theorem}.
We devide the proof into two steps.
The former part is to get the representation of the solution to (\ref{CP1}) by the wave packet transform.
%
In the latter part,
we estimate the each term in the representation generated in the former part
and prove Theorem \ref{main-theorem} by using the lemma in \cite{KKI}.

%
%
%
Let $t_0\in\mathbb{R}^n$ and $\varphi_0\in\mathcal{S}(\mathbb{R}^n)\setminus\{0\}$ be fixed.
The Cauchy problem (\ref{CP1}) is transformed by the wave packet transform with $\varphi(t,x)=e^{it\Delta/2}\varphi_0$ (defined in $\eqref{w-l}$) to
\begin{align}
\label{wpt-CP1}
\begin{cases}
\Big[i\partial_t+iA(t,x)\xi\cdot\nabla_x -i\displaystyle\Big(\frac{1}{2}\sum_{j,k=1}^n\nabla a_{j,k}(t,x)\xi_j\xi_k+\nabla_x V(t,x)\Big)\cdot\nabla_\xi\\
\hspace{6cm}+f(t,x,\xi)\Big]W_{\varphi(t)}u(t, x, \xi) = Ru(t, x, \xi), \\
W_{\varphi(0)}u(0, x, \xi)=W_{\varphi_0}u_0(x, \xi),
\end{cases}
\end{align}
where
\begin{align*}
f(t;x,\xi)=-\displaystyle\frac{1}{2}\sum_{j,k=1}^n(a_{j,k}(t,x)-x \cdot \nabla_{x}a_{j,k}(t,x))\xi_{j}\xi_{k}
+V(t,x)-x \cdot \nabla_x V(t,x)
\end{align*}
and
\begin{align*}
&Ru(t,x,\xi;\varphi_0)\\
&=\frac{1}{2}\sum_{j,k=1}^n\Bigg[(a_{j,k}(t,x)-\delta_{j,k})W_{\varphi^{(e_{j,k})}(t)}u(t,x,\xi)+\partial_{x_k}a_{j,k}
(t,x)W_{\varphi^{(e_j)}(t)}u(t,x,\xi)\\
&+\sum_{l=1}^n\left(\partial_{x_l} a_{j,k}(t,x)W_{\varphi^{(e_{j,k})}_{(e_l)}(t)}u(t,x,\xi)-2i\partial_{x_l} a_{j,k}(t,x)\xi_jW_{\varphi^{(e_{k})}_{(e_l)}(t)}u(t,x,\xi)\right)\\
&-i\partial_{x_k}a_{j,k}(t,x)\xi_jW_{\varphi(t)}u(t,x,\xi)+\dint u(t,y) (\partial_{y_k}\mathcal{A}^{j,k}_{M})(t,x,y)\partial_{y_j}(\overline{\varphi(y-x)}e^{-iy\xi})dy\\
&+\dint u(t,y)\mathcal{A}^{j,k}_{M}(t,x,y)\partial_{y_k}\partial_{y_j}(\overline{\varphi(y-x)}e^{-iy\xi})dy+\dint \overline{\varphi(t,y-x)}\mathcal{V}_{M}(t,x,y)u(t,y)e^{-iy\xi}dy
\Bigg],
\end{align*}
where $e_{j,k}=e_j+e_k$, $\mathcal{A}_M^{j,k}=\mathcal{A}_1^{j,k}+\mathcal{A}_2^{j,k}$, $\mathcal{V}_M=\mathcal{V}_1+\mathcal{V}_2$,
\begin{align*}
\mathcal{A}_1^{j,k}(t,x,y)&=\sum_{2\leq|\alpha|\leq M-1}\frac{1}{\alpha!}(\partial_x^\alpha a_{j,k})(t,x)(y-x)^\alpha,\\
\mathcal{A}_2^{j,k}(t,x,y)&=\sum_{|\alpha|=M}(M/\alpha!)\int_0^1(1-\theta)^{M-1}\partial_x^\alpha a_{j,k}(t,x+\theta(y-x))d\theta (y-x)^\alpha,\\
\mathcal{V}_1(t,x,y)&=\sum_{2\leq|\alpha|\leq M-1}\frac{1}{\alpha!}(\partial_x^\alpha V)(t,x)(y-x)^\alpha
\end{align*}
and
\begin{align*}
\mathcal{V}_2(t,x,y)=\sum_{|\alpha|=M}({M}/\alpha!)\int_0^1(1-\theta)^{M-1}\partial_x^\alpha V(t,x+\theta(y-x))d\theta (y-x)^\alpha.
\end{align*}

Here we use the Taylor expansion for $M\geq2$ as follows:
\begin{align*}
a_{j,k}(t,y)
=&a_{j,k}(t,x)+\nabla_x a_{j,k}(t,x)\cdot(y-x)+\mathcal{A}_M^{j,k}(t,x,y)\\
V(t,y)
=&V(t,x)+\nabla_x V(t,x)\cdot(y-x)+\mathcal{V}_M(t,x,y).
\end{align*}

%
%
Let $x(s;t_0,x,\xi)$ and  $\xi(s;t_0,x,\xi)$ be solutions of $\eqref{h-eq}$.
Then we have by the method of characteristic curve
\begin{align}\label{rep-by-wpt}
&W_{\varphi(t)}u(t,x(t;t_0,x,\xi),\xi(t;t_0,x,\xi))\\
\nonumber&=e^{i\int_0^tf(s)ds}W_{\varphi_0}u(0,x(0;t_0,x,\xi),\xi(0;t_0,x,\xi))\\
\nonumber&\hspace{3cm}-i\dint_0^te^{i\int_s^tf(\theta)d\theta}Ru(s,x(s;t_0,x,\xi),\xi(s;t_0,x,\xi);\varphi_0)ds
\end{align}
Substituting $\varphi_\lambda(-t_0)$ and $\lambda\xi$ for $\varphi_0$ and $\xi$ in (\ref{rep-by-wpt}), we obtain the following representation of solution to (\ref{CP1}).
\begin{align}\label{wpt of CP1}
&W_{\varphi_\lambda(t-t_0)}u(t,x(t;t_0,x,\lambda\xi),\xi(t;t_0,x,\lambda\xi))\\ 
\nonumber&=e^{i\int_0^tf(s)ds}W_{\varphi_\lambda(-t_0)}u(0,x(0;t_0,x,\lambda\xi),\xi(0;t_0,x,\lambda\xi))\\
\nonumber&\hspace{3cm}-i\dint_0^te^{i\int_s^tf(\theta)d\theta}R u(s,x(s;t_0,x,\lambda\xi),\xi(s;t_0,x,\lambda\xi);\varphi_\lambda(-t_0))ds.
\end{align}

%
%
In order to complete the proof, we use the following proposition.
\begin{Prop}[\cite{KKI}]
Let $u_0\in L^2(\mathbb{R}^n)$, $\varphi_0=e^{-|x|^2/2}$ and
$\varphi_\lambda(t)$ be satisfying $\eqref{w-l}$.
Then $(x_0,\xi_0)\notin WF(u_0)$ if and only if
there exist a neighborhood  $K$ of $x_0$ and a conic neighborhood $\Gamma$ of $\xi_0$ 
such that for all $N\in\mathbb{N}$, $a\geq1$ and $\varphi_0\in \mathcal{S}(\mathbb{R}^n)\setminus\{0\}$,
there exists $C_{N,a,\varphi_0}>0$ satisfying
\begin{align*}
|W_{\varphi_{\lambda}}u_0 (x, \lambda\xi)|\leq C_{N,a,\varphi_0}\lambda^{-N}
\end{align*}
for $\lambda\geq1$ and $(x,\xi)\in K\times\Gamma$ with $a^{-1}\leq|\xi|\leq a$.
\end{Prop}

We show only sufficiency. The other case can be shown in the same way.
We fix $a\geq1$. Let $K$ be a neighborhood of $x_0$ and $\Gamma$ be a conicneighborhood of $\xi_0$ satisfying (\ref{m-th})
for $t=t_0,\ \lambda\geq1,\ (x,\xi)\in K\times\Gamma$\ with\ $a^{-1}\leq|\xi|\leq a$.
%
%
Then we show the following claim\ $P(N,\varphi_0)$:\\
"There exists a constant $C_{N,a,\varphi_0}>0$ such that
\begin{align}
\label{ind-1}
|W_{\varphi_\lambda(t-t_0)}u(t,x(t;t_0,x,\lambda\xi),\xi(t;t_0,x,\lambda\xi))|\leq C_{N,a,\varphi} \lambda^{-N/2}
\end{align}
for $\lambda\geq1,\ (x,\xi)\in K\times\Gamma$ with $a^{-1}\leq|\xi|\leq a$, $0\leq t\leq t_0$".
We show that claim\ $P(N,\varphi_0)$ holds for all $N\in\mathbb{N}\cup\{0\}$ and $\varphi_0\in \mathcal{S}(\mathbb{R}^n)\setminus\{0\}$ by induction with respect to $N$.
First, we show $P(0,\varphi_0)$. For fixed $\varphi_0\in\mathcal{S}(\mathbb{R}^n)\setminus\{0\}$, we have
\begin{align*}
&|W_{\varphi_{\lambda}(t-t_0)}u(t, x(t; t_0, x, \lambda\xi), \xi(t; t_0, x, \lambda\xi))|\\
&\leq\dint|\varphi_\lambda(t-t_0, y-x)||u(t,y)|dy\\
&\leq\|\varphi_\lambda(t-t_0,\cdot)\|_{L^2}\|u(t,\cdot)\|_{L^2}\\
&=\|\varphi_0\|_{L^2}\|u_0\|_{L^2}.
\end{align*}
Hence, $P(0,\varphi_0)$ holds $N=0$ for all $\varphi_0\in\mathcal{S}(\mathbb{R}^n)\setminus\{0\}$.
Next, we fix $\varphi_0\in \mathcal{S}(\mathbb{R}^n)\setminus\{0\}$.
We show $P(N+1, \varphi_0)$ assuming $P(N, \varphi_0)$ for all $\varphi_0\in \mathcal{S}(\mathbb{R}^n)\setminus\{0\}$.
Then it suffices to prove the following statement:\\
``There exists a constant $C_{N,a,\varphi_0}>0$\ such that
$$|Ru(s, x(s; t_0, x, \lambda\xi), \xi(s; t_0, x, \lambda\xi);\varphi_\lambda(-t))|\leq C_{N,a,\varphi_0}\lambda^{-(N+1)/2}$$
for $\lambda\geq1,\ (x,\xi)\in K\times\Gamma$ with $a^{-1}\leq|\xi|\leq a,\ 0\leq s\leq t_0$,''
since the first term of the right hand side of $\eqref{wpt of CP1}$ is estimated by $\lambda^{-(N+1)/2}$ because of the condition of $u_0$.
%
%

We classify the term of $Ru$ as follows:
\begin{align*}
R_1(t,x,\xi;\varphi_0)=\frac12\sum_{j,k=1}^n\Bigg[&(a_{j,k}(t,x)-\delta_{j,k})W_{\varphi^{e_{j,k}}(t)}u(t,x,\xi)
-i\partial_{x_k}a_{j,k}\xi_jW_{\varphi(t)}u(t,x,\xi)\\
&-2i\sum_{l=1}^n\partial_{x_l}a_{j,k}(t,x)\xi_jW_{e^{it\Delta/2}(x_{l}\partial_{x_k}\varphi_0)}u(t,x,\xi)\\
&+\sum_{|\alpha|=2}\partial_{x}^\alpha a_{j,k}(t,x)\xi_j\xi_kW_{e^{it\Delta/2}(x^\alpha\varphi_0)}u(t,x,\xi)\Bigg]
\end{align*}
and
\begin{align*}
R_2(t,x,\xi;\varphi_0)&=Ru(t,x,\xi;\varphi_0)-R_1(t,x,\xi;\varphi_0).
\end{align*}
$R_2(t,x,\lambda\xi;\varphi_\lambda(-t))$ can be estimated in the same way in \cite{KaI} by using the equality
\begin{align*}
x^\beta\partial_x^\alpha(\varphi_\lambda(t,x))
=\lambda^{|\alpha|/2}((\lambda^{-1/2}x+it\lambda^{1/2}\nabla)^\beta(\partial_x^{\alpha}\varphi_0))_{\lambda}(t,x,\xi),
\end{align*}
where $(x_j\varphi_0)_{\lambda}(t,x)$ and $(\partial_{x_j}\varphi_0)_{\lambda}(t,x)$ are denoted
by $\eqref{w-l}$ with $\varphi_0\rightarrow x_j\varphi_0$ and $\varphi_0\rightarrow \partial_{x_j}\varphi_0$ for $j=1,\ldots,n$, respectively.


%
%
Next, we shall  estimate $R_1(t,x,\lambda\xi;\varphi_\lambda(-t))$.
%
By Proposition \ref{c-o-prop}, we have
$|\nabla_x a_{j,k}(s,x(s))|\leq C(1+\lambda|s-t_0|)^{-\rho-1}$ and $|\xi_j(s)|\leq C\lambda$ for large $\lambda$ and $j,k=1,\ldots,n$,
which yields that
\begin{align*}
&\Big|\dint_0^t \partial_{x_l} a_{j,k}(s,x(s)) \xi_j(s)W_{(y_l\partial_{y_k}\varphi)_\lambda(s-t_0)}u(s,x(s;t_0,x,\lambda\xi),\xi(s;t_0,x,\lambda\xi))ds\Big|\\
&\leq C\lambda\dint_0^\infty (1+\lambda|s-t_0|)^{-\rho-1} ds\times \displaystyle \sup_{\substack{0\leq t\leq t_0,x\in K,\\ \xi\in\Gamma,a^{-1}\leq|\xi|\leq a}} |W_{(y_l\partial_{y_k}\varphi)_\lambda(t-t_0)}u(s,x(t),\xi(t))|\\
&\leq C\displaystyle \sup_{t,x,\xi} |W_{(y_l\partial_{y_k}\varphi)_\lambda(t-t_0)}u(s,x(t),\xi(t))|
\end{align*}
for $l=1,\ldots,n$.
Let $b>0$ be fixed, take $\kappa$ and $m$ satisfying $(\rho-1)^{-1}<\kappa<m-1$  and define $\Phi(t)$ as follows:
\begin{align}\label{def-ph}
\Phi(t)=\Phi(t;\lambda,x,\xi)\equiv\displaystyle\sum_{l=0}^\infty\frac{{C_m(t)}^{l+1}}{l!}\sum_{|\alpha+\beta|=2l}\Big\langle W_{e^{\frac{i}{2}t\Delta}(x^\alpha\partial_x^\beta\varphi_0)_\lambda}u(t, x, \xi)\Big\rangle_{N,\lambda},
\end{align}
where $\langle\cdot\rangle_{N,\lambda}=\sqrt{\lambda^{-Nb}+|\cdot|^2}$ and
$C_m(t)=e^{\kappa(1+\lambda t)^{-\rho+1}-m}$.
If $\varphi_0(x)=\sum_{\gamma}C_{\gamma}x^{\gamma}e^{-|x|^2}$, the above sum converges absolutely and uniformly with respect to $t$.
Indeed,
for $\varphi_0=e^{-|x|^2}$,
we get $|x^\alpha\partial_x^\beta\varphi_0|\leq 2^{2l}\max\{1,|x|^{2l}\}\varphi_0\leq2^{2l}(l+1)!(1+x^2)^{-1}$ for any $\alpha,\, \beta\in \mathbb{Z}_+^n$ with $|\alpha+\beta|=2l$ by the inequality $e^{x^2}\geq (1+x^{2l+2})/(l+1)!$.
Thus there exists the positive constant $M$ independent of $l$ such that $\langle W_{e^{\frac{i}{2}t\Delta}(x^\alpha\partial_x^\beta\varphi_0)_\lambda}u(t, x, \xi)\rangle_{N,\lambda}\leq M2^{l}(l+1)!$,
which implies the convergence of the above sum for large $\lambda$.
The case for general $\varphi_0$ can be proven similarly.

For $l\in\mathbb{N}\cup\{0\}$ and $\alpha, \beta\in \mathbb{Z}_+^n$ with $|\alpha+\beta|=2l$, the inequality
$$|W_{e^{it\Delta/2}(y^\alpha\partial_y^\beta\varphi_0)_\lambda} u(t, x(t), \xi(t))|\leq C\Phi(t;\lambda,x(t),\xi(t))$$
holds for any $t\in[0,t_0]$ and $(x,\xi)\in K\times\Gamma$ with $a^{-1}\leq|\xi|\leq a$,
where $C$ depends  only on $n$, $t_0>0$,  $M$ and $N$.
Thus it suffices to estimate $\Phi(t)$.
Since $(d/dt)\langle z(t)\rangle\leq|(d/dt)z(t)|$ holds for complex-valued function $z(t)$,
we have
\begin{align*}
\displaystyle\frac{d}{dt}\Big\langle W_{e^{it\Delta/2}(x^\alpha\partial_x^\beta\varphi_0)_\lambda}&u(t, x(t), \xi(t))\Big\rangle_{N,\lambda}
=\frac{d}{dt}\Big\langle e^{i\int_0^tf(s)ds}W_{e^{it\Delta/2}(x^\alpha\partial_x^\beta\varphi_0)_\lambda}u(t, x(t), \xi(t))\Big\rangle_{N,\lambda}\\
&\leq\left|\frac{d}{dt}(e^{i\int_0^tf(s)ds}W_{e^{it\Delta/2}(x^\alpha\partial_x^\beta\varphi_0)_\lambda}u(t, x(t), \xi(t)))\right|\\
&=\left|e^{i\int_0^tf(s)ds}\Big(i\partial_t+ix(t)\cdot\nabla_x+i\xi(t)\cdot\nabla_\xi
+f(t,x,\xi)\Big)W_{\varphi(t)}u(t, x, \xi) )\right|\\
&=|Ru(t, x(t), \xi(t))|,
\end{align*}
which and the equality that
\begin{align*}
\Phi^{'}(t)
&=\displaystyle\sum_{l=0}^\infty\frac{(l+1){C_m(t)}^{l}C'_m(t)}{l!}\sum_{|\alpha+\beta|=2l}\Big\langle W_{e^{it\Delta/2}(x^\alpha\partial_x^\beta\varphi_0)_\lambda}u(t, x(t), \xi(t))\Big\rangle_{N,\lambda}\\
&+\displaystyle\sum_{l=0}^\infty\frac{{C_m(t)}^{l+1}}{l!}\sum_{|\alpha+\beta|=2l}\frac{d}{dt}\Big\langle W_{e^{it\Delta/2}(x^\alpha\partial_x^\beta\varphi_0)_\lambda}u(t, x(t), \xi(t))\Big\rangle_{N,\lambda}
\end{align*}
yield that
\begin{align*}
\Phi'(t)&\leq\displaystyle\sum_{l=0}^\infty\frac{(l+1){C_m(t)}^{l}C'_m(t)}{l!}\sum_{|\alpha+\beta|=2l}\Big\langle W_{e^{it\Delta/2}(x^\alpha\partial_x^\beta\varphi_0)_\lambda}u(t, x(t), \xi(t))\Big\rangle_{N,\lambda}\\
&+\displaystyle\sum_{l=0}^\infty\frac{{C_m(t)}^{l+1}}{l!}\sum_{|\alpha+\beta|=2l}|Ru(t, x(t), \xi(t))|\\
&\leq \lambda\kappa (1+\lambda t)^{-\rho}(1-\rho)\sum_{l=0}^\infty\frac{(l+1){C_m(t)}^{l}C_m(t)}{(l+1)!}\sum_{|\alpha+\beta|=2l}\Big\langle W_{e^{it\Delta/2}(x^\alpha\partial_x^\beta\varphi_0)_\lambda}u(t, x(t), \xi(t))\Big\rangle_{N,\lambda}\\
&+\Big(C\lambda(1+\lambda t)^{-\rho}\sum_{l=0}^\infty\frac{{C_m(t)}^{l}}{l!}\sum_{|\alpha+\beta|=2l}\Big\langle W_{e^{it\Delta/2}(x^\alpha\partial_x^\beta\varphi_0)_\lambda}u(t, x(t), \xi(t))\Big\rangle_{N,\lambda}\Big)+C\lambda^{-(N+1)/2}\\
&=C(1-(\rho-1)\kappa)\frac{\lambda}{(1+\lambda t)^{\rho}}\sum_{l=0}^\infty\frac{{C_m(t)}^{l}}{l!}\sum_{|\alpha+\beta|=2l}\Big\langle W_{e^{it\Delta/2}(x^\alpha\partial_x^\beta\varphi_0)_\lambda}u(t, x(t), \xi(t))\Big\rangle_{N,\lambda}\\
&+C\lambda^{-(N+1)/2}\\
\end{align*}
The choice of $\kappa$ implies that $\Phi'(t)\leq C\lambda^{-(N+1)/2}$.
Hence, we have $\Phi(t)\leq \Phi(0)+tC^{'}\lambda^{-(N+1)/2}\leq \Phi(0)+t_0 C^{'}\lambda^{-(N+1)/2}$.
In particular, we have $\Phi(0)\leq C\lambda^{-(N+1)/2}$.
The other terms in $R_1(t,x,\lambda\xi;\varphi_\lambda(-t))$ can be estimated similarly.
Therefore, we complete the proof of Theorem \ref{main-theorem}.
\section{Proof of Collorary \ref{Cor2}}
In this section, we give a proof of Collorary \ref{Cor2}.
The differences between $x(0;t,x,\lambda\xi)$ and  $x-t\lambda\xi$ 
between and $\xi(0;t,x,\lambda\xi)$ and $\lambda\xi$
is small for large $\lambda$.
Using this property, we prove Collorary \ref{Cor2}.
\begin{proof}[Proof of Corollary \ref{Cor2}]
In order to complete the proof, it suffices to prove that for any neighborhood $K_0$ of $x_0$ and conic neighborhood $\Gamma_0$ of $\xi_0$,
there exist neighborhoods ${K}_1$, ${K}_2$ of $x_0$ and  conic neighborhoods ${\Gamma}_1$, ${\Gamma}_2$ of $\xi_0$
such that for large $\lambda$
\begin{align}
\label{subs}
\left\{(x-t\lambda\xi,\lambda\xi)\,\middle|\,(x,\xi)\in{K}_1\times{\Gamma}_1\right\}
\subset
\left\{(x(0;t,x,\lambda\xi),\xi(0;t,x,\lambda\xi))\,\middle|\,(x,\xi)\in{K}_0\times{\Gamma}_0\right\}
\end{align}
and
\begin{align}
\label{sups}
\left\{(x(0;t,x,\lambda\xi),\xi(0;t,x,\lambda\xi))\,\middle|\,(x,\xi)\in{K}_2\times{\Gamma}_2\right\}
\subset
\left\{(x-t\lambda\xi,\lambda\xi)\,\middle|\,(x,\xi)\in{K}_0\times{\Gamma}_0\right\}.
\end{align}
Indeed, $\eqref{subs}$ implies
\begin{align}
\label{subs1}
\sup_{(x,\xi)\in V_1}|W_{\varphi_{\lambda}{(-t)}}u_0 (x-t\lambda\xi,\lambda\xi)|
\leq\sup_{(x,\xi)\in V_0}|W_{\varphi_{\lambda}{(-t)}}u_0 (x(0,t,x,\lambda\xi), \xi(0,t,x,\lambda\xi))|
\end{align}
and $\eqref{sups}$ implies
\begin{align}
\label{sups1}
\sup_{(x,\xi)\in V_2}|W_{\varphi_{\lambda}{(-t)}}u_0 (x(0,t,x,\lambda\xi),\xi(0,t,x,\lambda\xi))|
\leq\sup_{(x,\xi)\in V_0}|W_{\varphi_{\lambda}{(-t)}}u_0 (x-t\lambda\xi, \lambda\xi)|,
\end{align}
which yields that Collorary \ref{Cor2} follows from Theorem \ref{main-theorem},
where $V_j=\{(x,\xi)|x\in K_j,\xi\in \Gamma_j,a^{-1}\leq|\xi|\leq a\}$ for $j=0,1,2$.

Without loss of generality, we have $K_0=B_r(x_0)$ and $\Gamma_0=\{\xi\in\mathbb{R}^n\,|\,(\xi\cdot\xi_0)/|\xi||\xi_0|>1-\gamma\}$
for some $r>0$ and $0<\gamma<1$.
We shall only prove $\eqref{subs}$ since $\eqref{sups}$ can be proven similarly.
Then there exists $\lambda_1\geq1$ such that
\begin{align*}
\delta_1(\lambda)\equiv|x(0;t,x,\lambda\xi)-(x-t\lambda\xi)|\leq r/2
\end{align*}
and
\begin{align*}
\delta_2(\lambda)\equiv|\xi(0;t,x,\lambda\xi)-\lambda\xi|\leq |\xi_0|\sqrt{\gamma/2}
\end{align*}
for any $(x,\xi)\in V_0$ and $\lambda\geq\lambda_1$,
since $\delta_1(\lambda),\delta_2(\lambda)=\mathcal{O}(\lambda^{-\rho+1})$ by Lemma \ref{c-l-2}.
Taking $K_1=B_{r/2}(x_0)$ and  $\Gamma_1=\{\xi\in\mathbb{R}^n\,|\,|\xi\cdot\xi_0|/|\xi||\xi_0|>\sqrt{(2-\gamma)/2}\}$, we have
\begin{align}
\label{subs2}
\left\{(x-t\lambda\xi,\lambda\xi)\,\middle|\,(x,\xi)\in{K}_1\times{\Gamma}_1\right\}
\subset
\left\{(x+\delta_1(\lambda),\xi+\delta_2(\lambda))\,\middle|\,(x,\xi)\in{K}_0\times{\Gamma}_0\right\}
\end{align}
for $\lambda\geq\lambda_1$.
Since $\Gamma_1\subset \Gamma_0$, $\eqref{subs2}$ shows $\eqref{subs}$.
\end{proof}


\begin{thebibliography}{99}
\bibitem{CF}A. C\'ordoba and C. Fefferman, Wave packets and Fourier integral operators,
{\it Comm. Partial Differential Equations}
{\bf 3} (1978), 979--1005.
\bibitem{Fo}G. B. Folland, Harmonic analysis in phase space, Ann. of Math. Studies No.122, Princeton Univ. Press, Princeton, NJ, (1989).
\bibitem{HW}A. Hassell and J. Wunsch, On the structure of the Schr\"odinger propagator,
{\it  in Partial differential equations and inverse problems, Contemp. Math., Amer. Math. Soc., Providence, RI}
{\bf 362} (2004), 199--209.
\bibitem{Hor}L. H\"ormander, Fourier integral operators, I, Acta Math. 127 (1971), 79–183.
\bibitem{KKI}K. Kato, M. Kobayashi and S. Ito, Remark on characterization of wave front set by wave packet transform, arXiv:1408.1370v1.
\bibitem{KaI}K. Kato, and S. Ito, Singularities for solutions to time dependent Schr\"odinger equations with sub-quadratic potential, SUT J. Math. Vol. 50 No.2 (2014), 383--398.
\bibitem{Me}R. Melrose, Spectral and scattering theory for the Laplacian on asymptotically Euclidian spaces,
Spectral and scattering theory (Sanda, 1992)
{\it Lecture Notes in Pure and Appl. Math.}
{\bf 161} (1994), 85--130.
\bibitem{Na}S. Nakamura, Wave front set for solutions to Schr\"odinger equations, J. Funct. Anal. 256 (2009), 1299--1309.
\bibitem{Oo}T. $\overline{\text{O}}$kaji, A note on the wave packet transforms, Tsukuba J. Math. 25 (2001), 383--397.
\end{thebibliography}
\end{document}